# Digital Twin for Estimating QoT Statistics in Presence of PDL and Transceiver Imperfections

Ambashri Purkayastha[(1)], Camille Delezoide[(1)], Vinod Bajaj[(1)], Mounia Lourdiane[(2)], Cédric Ware[(3)], Patricia Layec[(1)]

[(1)] Nokia Bell Labs, France, ambashri.purkayastha@nokia.com
[(2)] SAMOVAR, Télécom SudParis, Institut Polytechnique de Paris, France
[(3)] LTCI, Télécom Paris, Institut Polytechnique de Paris, France

**Abstract** *We propose a physics-based digital twin to predict the statistical QoT distribution of a realistic optical lightpath. We demonstrate up to 0.73 dB accuracy improvement in worst-case SNR prediction for short distance transmissions in linear regime.* ©2025 The Author(s)

**Introduction**

Accurately predicting the quality of transmission (QoT) is a major issue in optical networks. Beyond traditional models integrating margins for worst-case design [1], digital twins (DTs) are meant to precisely reproduce the QoT of individual lightpaths. The main targeted application is sandboxing, allowing faithful simulation of virtual configurations before implementation [2].

In a high-fidelity twin, the inner QoT model ideally reproduces the time-dependence of the twinned lightpath. This is partially achieved by leveraging measurements on model input parameters rather than fixed values. Yet, stochastic phenomena such as polarization dependent loss (PDL) inherently limit the QoT prediction accuracy at any time. In practice we thus need to predict the statistical properties of the QoT rather than its instantaneous values. Those are later used for instance to determine adequate margins through the analysis of the lower tail [3].

The prediction of the QoT statistics has been extensively studied in long reach transmissions where the dominant noise sources are the amplified spontaneous emission (ASE) noise from amplifiers and the nonlinear impairments [3,4]. As coherent transmissions are increasingly deployed on shorter distances using low-cost transceivers (TRx), we propose in this paper a DT-based statistical QoT model which accounts for both PDL and polarization dependent TRx impairments, thereby compatible with transmissions limited by transceiver noises [5-7].

First, we experimentally investigate the polarization-dependence of TRx noises for various central frequencies and propose a model to account for such imperfections. We then propose a DT-based workflow that leverages the proposed model to predict QoT statistics in the presence of PDL. Finally, we perform simulations of a PDL chain combined with calibrated TRx imperfections to test the DT's ability to satisfactorily predict the QoT's PDF and discuss the potential impact when predicting failure probabilities.

**QoT Model and Experimental Calibration**

The QoT of a lightpath is typically characterized by its SNR, i.e. the signal-to-noise ratio of the signal before decision, averaged between both polarization tributaries, typically expressed as [8]:

$$\text{SNR}^{-1} = \alpha \cdot \text{GOSNR}^{-(1+\varepsilon)} + \text{SNR}_{\text{trx}}^{-1} \qquad (1)$$

where:
- The generalized optical signal-to-noise ratio (GOSNR) characterizes the line by combining ASE and nonlinear noise contributions
- $\text{SNR}_{\text{trx}}$ represents the TRx noises
- $\alpha$ accounts for the actual receiver bandwidth
- $\varepsilon$ accounts for the multiplicative noise

In what we refer to as the present mode of operation (PMO), the SNR of a lightpath is determined by first evaluating the GOSNR value using for instance the IGN model [9], then by applying calibrated values for $\text{SNR}_{\text{tr}}$, $\alpha$ and $\varepsilon$ to Eq. (1). Those values can be obtained using the experimental setup shown in Fig. 1, by operating EDFA in constant power mode. By varying the attenuation through the VOA, we obtain for a 32 GBaud DP-QPSK the SNR-GOSNR curve represented in Fig. 2(a) with the displayed fitted values.

In the PMO, it is assumed that the pre-FEC BER directly derives from $\langle SNR \rangle = (SNR_x + SNR_y)/2$, as:

$$BER \approx BER_{PMO} \equiv \frac{1}{b}\text{erfc}\left(\sqrt{\frac{\langle SNR \rangle}{2c}}\right) \qquad (2)$$

where b and c depend on the modulation format [10]. However, when $\text{SNR}_x$ and $\text{SNR}_y$ significantly differ, this approximation leads to major errors in pre-FEC BER. The true pre-FEC BER value

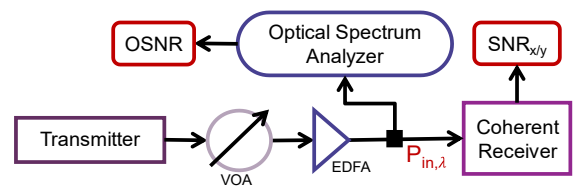

Fig.1: Experimental setup to calibrate the transceiver

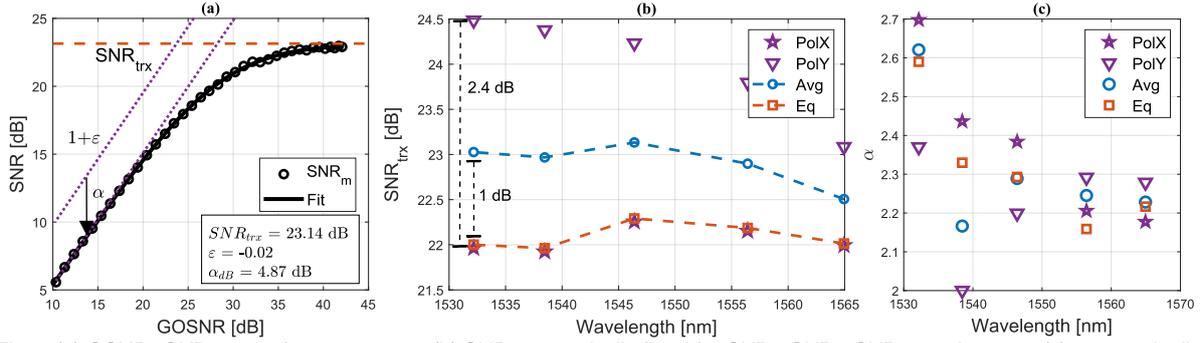

Fig.2: (a) OSNR-⟨SNR⟩ curve for at 1546 nm; (b) SNR$_{trx}$ numerically fitted for SNR$_x$, SNR$_y$, SNR$_{eq}$, and ⟨SNR⟩ ; (c) $\alpha$ numerically fitted for SNR$_x$, SNR$_y$, SNR$_{eq}$, and ⟨SNR⟩ , assuming $\varepsilon = 0$.

derives from the equivalent SNR, SNR$_{eq}$, as [11]:

$$BER \equiv \frac{BER_x + BER_y}{2} \equiv \frac{1}{b}\mathrm{erfc}\left(\sqrt{\frac{SNR_{eq}}{2c}}\right) \quad (3)$$

In the following, the QoT estimation error under the PMO is thus evaluated as $SNR_{eq} - \langle SNR \rangle$.

With the same experimental setup described in Fig. 1, we performed back-to-back calibrations to measure the SNR$_x$-OSNR and SNR$_y$-OSNR curves for five channel wavelengths covering the C-band. From the numerical fits of the ten curves, we obtain the polarization-dependent fitted values for SNR$_{trx}$ and $\alpha$ represented in Fig. 2(b-c). Note that since ε values are negligible for this transmission, cf. Fig. 2(a), we assumed $\varepsilon = 0$.

In the worst case observed in Fig. 2(b), at 1532 nm, the difference between the $SNR_{tr,x}$ and $SNR_{tr,y}$ is 2.4 dB, resulting in a 1 dB QoT estimation error under PMO at high GOSNR values, where TRx noises dominate. Additionally, experiments show a significant impact of the channel wavelength on the polarization-dependence of TRx imperfections. In Fig. 2(c), we further observe significant discrepancies in fitted $\alpha$ values depending on the polarization tributary, representing up to 12.5% of the average value.

These measurements demonstrate the importance of polarization-dependent TRx noises calibration and their integration in the QoT estimation model. To do so, we propose to express the SNR$_{x/y}$ as:

$$\mathrm{SNR}_{x/y} = \mathcal{G}(\mathrm{GOSNR}_{x/y}, \mathrm{SNR}_{tr,x/y}, \alpha_{x/y}, \varepsilon_{x/y}) \quad (4)$$

where $\mathcal{G}$ represents the same function as in Eq. (1) while $\alpha_{x/y}$, $\varepsilon_{x/y}$ and SNR$_{tr,x/y}$ are the polarization dependent parameters. To account for random QoT variations, we introduce GOSNR$_{x/y}$ as the per-polarization GOSNR in presence of PDL, verifying $GOSNR_{x/y} = GOSNR_0 +/- \delta$, GOSNR$_0$ being the mean GOSNR while $\delta$ accounts for PDL induced random QoT variations. Those depends on the magnitude, location and number of PDL elements. SNR$_{eq}$ is deduced from Eq. (3) through the BER$_{x/y}$ corresponding to SNR$_{x/y}$ [11].

**Proposed Lightpath Digital Twin**

Our proposed DT has a similar architecture to that defined in [5], but with a new workflow to account of the PDL and polarization-dependent TRx imperfections. It operates in two parallel modes: training and sandboxing.

In training mode, the DT continuously collects SNR$_{x/y}$ measurements from the coherent receiver to learn GOSNR$_{x/y}$ statistics. GOSNR$_{x/y}$ values are deduced from SNR$_{x/y}$ values inverting Eq. (4), then used to evaluate the PDF of $\delta$ in dB which is fitted by a lognormal distribution to finally obtain its mean $\mu_t$ and standard deviation $\sigma_t$.

In sandboxing mode, the updated values of $\mu_t$ and $\sigma_t$ are leveraged to predict QoT statistics for an alternative configuration of the same lightpath. This is achieved through the following steps:

1. Prediction of $GOSNR'_0$ as the new GOSNR$_0$ value in the alternative configuration.
2. Calculation of the predicted mean for $\delta$ as $\mu_p = \ln(GOSNR'_0) - \sigma_t^2/2$. We assume the standard deviation is unchanged: $\sigma_p = \sigma_t$.
3. Generation of $GOSNR'_x$ values following a lognormal distribution with $\mu_p$ and $\sigma_p$.
4. Calculation of associated $GOSNR'_y$ values as $GOSNR'_y = 2 \cdot GOSNR'_0 - GOSNR'_x$.
5. Calculation of $SNR'_{x/y}$ then SNR$_{eq}$ values using calibrated tributary dependent TRx parameters and generated $GOSNR'_{x/y}$ values.
6. Deduction of $f(SNR'_{eq,DT})$, the PDF of the SNR$_{eq}$ distribution predicted by the DT for the alternative configuration.

We performed numerical simulations using the PDL model described in [3] to validate the proposed workflow and outline differences between the DT and the PMO. We emulated a lightpath in linear regime with 20 spans, where each EDFA has a PDL of 0.4 dB in magnitude. Various GOSNR$_0$ values are obtained by uniformly varying EDFA gains. For each gain setting, we generated 10$^6$ $GOSNR'_{x/y}$ values corresponding to randomized PDL orientations following a uniform distribution [12].

In Fig. 3(a) we show the simulated time series for $SNR_{x/y}$, $SNR_{eq}$, and $\langle SNR \rangle$ for two $GOSNR_0$ values. At higher GOSNR on Fig. 3(a.1), the SNRs are dominantly impacted by TRx noises including polarization-dependence. This results in average offset of 1.3 dB offsets between $SNR_x$ and $SNR_y$ values. This leads to a mean QoT estimation error of 0.6 dB under PMO approximation, at it only accounts for $\langle SNR \rangle$. For sandboxing, at lower GOSNR on Fig. 3(a.2), we observe as expected that $SNR_x$ and $SNR_y$ are closer in comparison due to the relative reduction of the contribution of TRx noises in the SNRs, leading to a reduced QoT error under PMO's approximation. Since GOSNR decreases with transmission distance, it illustrates how the PMO holds for long reaches but increasingly fails for shorter ones.

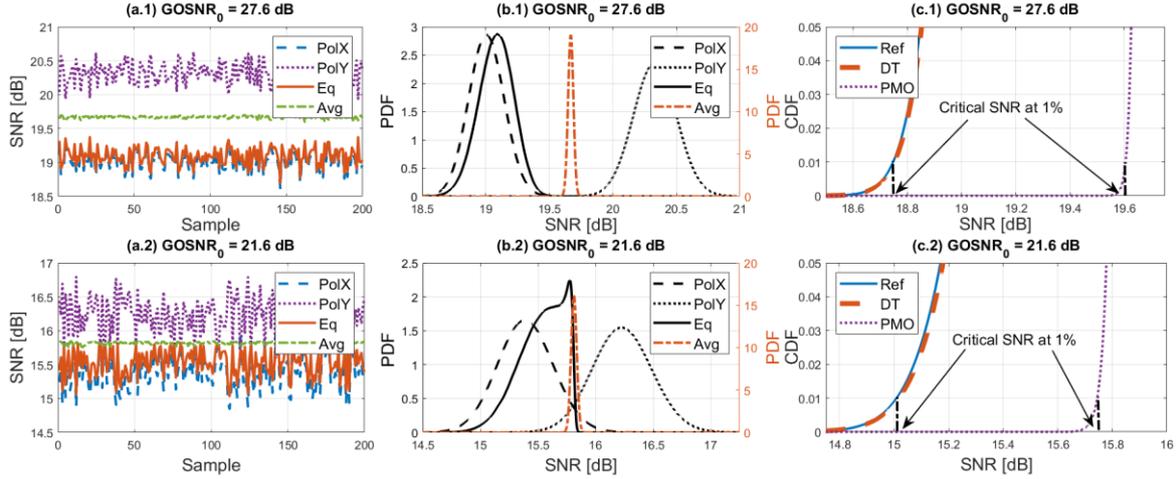

Fig. 3: (a) Time series evolution of SNRs at (1) training GOSNR, (2) sandboxing with higher ASE noise power (b) Corresponding PDFs of SNR distributions; (c) 5% of CDF of the QoT at the predicted by reference, DT and PMO at with critical SNR at 1%.

In Fig. 3(b), we plot the PDFs corresponding to the time series of Fig. 3(a). We first observe that the PDL model used for simulations leads to a narrow distribution for $\langle SNR \rangle$, consistent with the approximation of a constant value for $GOSNR_0$ made in step 4 of the sandboxing mode. We further observe for the higher GOSNR that the distribution of $SNR_{eq}$ is very close to that of $SNR_x$. Indeed, because of the large offset in average between $SNR_x$ and $SNR_y$ values, $SNR_{eq}$ values are mainly driven by the SNR of the worst-performing tributary, i.e. by $SNR_y$. In contrast, at the lower GOSNR value, the reduced offset between $SNR_x$ and $SNR_y$ leads to a distorted $SNR_{eq}$ distribution. A major use case of the proposed DT is to predict—from the $SNR_{eq}$ distribution—a critical SNR value, $SNR_n$ such that $P(SNR_{eq} < SNR_n) = n\%$. This comes down to determining a statistical worst-case QoT and using it to appropriately set margins. Accordingly, here we quantify our DT's performance from errors made on $SNR_n$ estimation. Training on $10^3$ points for $GOSNR_0$ = 27.6 dB, we predict the performance at $GOSNR_0'$ = 21.6 dB.

The predicted $SNR_n$ values can be directly read for arbitrary n values from the cumulative density function (CDF) of the QoT distribution. Thus, we plot in Fig. 3(c) both the CDF of $SNR_{eq}$ predicted by the DT and the CDF of $\langle SNR \rangle$ that would be considered under the PMO. We further plot the reference CDF obtained when performing the simulations at the current $GOSNR_0$ value. We directly observe that for both tested configurations, the CDFs predicted by the DT almost perfectly match the reference CDFs while the CDF corresponding to the PMO significantly overestimate the QoT. For n=1% at $GOSNR_0$ = 21.6 dB, this results in an error on $SNR_1$ of 0.74 dB for the PMO against only 0.004 dB for the DT. Also note that the DT outperforms the PMO to a greater extent for smaller n values.

**Conclusions**

We have proposed a digital twin for the prediction of QoT statistics for a realistic optical lightpath impacted by both PDL and polarization-dependent transceiver imperfections. We presented experimental results outlining polarization dependent transceiver noise discrepancies and proposed a model extension to account for it while also accounting for inline PDL. Through numerical simulations in linear propagation regime, we demonstrated that this model extension combined with the proposed DT workflow achieves significant improvements compared to the PMO when predicting critical SNR values corresponding to statistical worst-cases, up to 0.73 dB. The proposed twin would thus be particularly beneficial for predicting lower tail QoT statistics for cost effective short-reach polarization-multiplexed coherent transmissions.